\documentclass{article}

\PassOptionsToPackage{numbers, compress}{natbib}

\usepackage[preprint]{neurips_2023}

\usepackage[utf8]{inputenc} 
\usepackage[T1]{fontenc}    
\usepackage{hyperref}       
\usepackage{url}            
\usepackage{booktabs}       
\usepackage{amsfonts}       
\usepackage{nicefrac}       
\usepackage{microtype}      
\usepackage{xcolor}         

\usepackage{times}
\usepackage{latexsym}
\usepackage{enumitem}

\usepackage{graphicx}
\usepackage{amssymb}
\usepackage{comment}

\usepackage{caption}
\usepackage{subcaption}
\usepackage{booktabs}

\usepackage{multicol}
\usepackage{multirow}
\usepackage{array, tabularx,  ragged2e,  booktabs}
\usepackage{natbib}

\usepackage[utf8]{inputenc}
\usepackage{graphicx}
\usepackage{pgfplots}
\pgfplotsset{compat=1.14}

\usepackage{ctable}
\usepackage{wrapfig}
\usepackage{tikz}
\usetikzlibrary{positioning,arrows,trees,shapes,fit,shadows}

\usepackage{amsfonts}
\usepackage{ifthen}
\usepackage{booktabs}
\usepackage{amssymb}
\usepackage{supertabular}
\usepackage{array}
\usepackage{multirow}
\usepackage{fancyhdr}
\usepackage[normalem]{ulem}

\usepackage{amssymb}
\usepackage{supertabular}
\usepackage{array}
\usepackage{multirow}
\usepackage{fancyhdr}
\usepackage{psfrag}
\usepackage{amsmath}
\usepackage{hhline}
\usepackage{type1cm}
\usepackage{lettrine}
\usepackage[colorinlistoftodos,prependcaption,textsize=tiny]{todonotes}

\usepackage{algorithm}
\usepackage[noend]{algpseudocode}

\usepackage{xspace}

\usepackage{babel}
\usepackage[export]{adjustbox}
\usepackage{cleveref}

\usepackage{pifont}

\usepackage{scrextend}

\crefformat{section}{\S#2#1#3} 
\crefformat{subsection}{\S#2#1#3}
\crefformat{subsubsection}{\S#2#1#3}

\usepackage{amsmath,amsfonts,bm}

\def\eqref#1{equation~\ref{#1}}

\def\1{\bm{1}}

\makeatletter
\newcommand{\xRightarrow}[2][]{\ext@arrow 0359\Rightarrowfill@{#1}{#2}}
\makeatother

\DeclareMathAlphabet{\mathsfit}{\encodingdefault}{\sfdefault}{m}{sl}
\SetMathAlphabet{\mathsfit}{bold}{\encodingdefault}{\sfdefault}{bx}{n}

\newcommand{\Ni}{({\em i})~}
\newcommand{\Nii}{({\em ii})~}
\newcommand{\Niii}{({\em iii})~}

\usepackage{float}

\DeclareCaptionLabelFormat{andtable}{#1~#2  \&  \tablename~\thetable}

\usepackage{tikz-cd}
\usepackage{mathtools}
\usepackage{todonotes}
\usepackage{xcolor}

\definecolor{darkblue}{rgb}{0, 0, 0.5}
\hypersetup{
    colorlinks=true,
    linkcolor=darkblue,
    citecolor=darkblue,
    pdfpagemode=FullScreen,
    }

\title{Mixture-of-Parallelisms: Towards Memory-Efficient Training Stack for Mixture-of-Experts Models}

\author{%
    Xuan-Phi Nguyen$^{\dagger}$\thanks{Project lead \& corresponding authors: \href{mailto:xnguyen@salesforce.com}{\{xnguyen,sjoty\}@salesforce.com}} \qquad Shrey Pandit$^{\dagger}$ \qquad Yiran Zhao$^{\dagger}$
    \And
    Semih Yavuz \qquad Silvio Savarese \qquad Shafiq Joty$^{*}$ \\
    Salesforce AI Research
}

\begin{document}

\maketitle

\setcounter{footnote}{0} 

\begin{abstract}

This paper proposes a memory-efficient training stack for Mixture-of-Experts (MoE) models. It is a training paradigm that combines and specializes various existing and novel parallelism techniques at different layers and stages of the Mixture-of-Experts (MoE) model training pipeline. It leverages these techniques to achieve maximal efficiency given the physical constraints of CPU, CPU memory, GPU HBM memory, and the CPU-GPU, GPU-GPU, and node-node communication bandwidth of the GPU cluster.
It also contains a novel strategy for the optimizer step to achieve high throughput and memory efficiency, enabling practitioners to conduct lossless pre-training/fine-tuning of trillion-parameter scale models, at a million context length, with just under 12 8x H200 GPU nodes, with state-of-the-art throughput and memory efficiency.
In our experiments, MoP delivers $4.7\text{--}8.2\times$ higher per-GPU throughput than a strongly-tuned FSDP2 baseline (with the gap widening at larger scale) and sustains training at context lengths up to 1M tokens, where the baseline runs out of memory beyond 64--128K.

\end{abstract}

\section{Introduction}\label{sec:intro}

Sparse Mixture-of-Experts (MoE) architectures have become the dominant recipe for
scaling language models, because they decouple parameter count from per-token
compute: routing each token to only $k$ of $E$ experts lets the parameter budget---and
with it model quality---grow far faster than the FLOPs per token
\citep{gshard_lepikhin2020,fedus2022switch}. The frontier this enables,
trillion-parameter models trained at near-million-token context, is no longer
compute-bound on modern accelerators; it is bound by \emph{GPU memory} and by the
\emph{communication bandwidth} of the cluster. A single trillion-parameter model needs
$\sim\!16\Theta$ bytes of persistent optimizer-related state alone
\citep{loshchilov2019decoupled} (\S\ref{sec:bg:memory}), tens of terabytes that no
single device can hold, while long-context activations add a transient footprint that
grows with sequence length and depth. Training such models therefore hinges on how
effectively a system spreads state and traffic across the memory hierarchy and
interconnects of the machine.

\paragraph{The limitation of one-size-fits-all parallelism.} The prevailing systems
attack this problem with a \emph{single global plan applied uniformly to every layer}.
Megatron-LM \citep{shoeybi2019megatron,korthikanti2023reducing} composes tensor,
pipeline, and sequence parallelism into a 3D grid $W\!=\!d\,t\,p$; ZeRO and its PyTorch
realization FSDP \citep{rajbhandari2020zero,zhao2023fsdp} shard the persistent state
across all data-parallel ranks; expert parallelism \citep{rajbhandari2022deepspeedmoe}
partitions experts across devices. Each is excellent at the bottleneck it targets and
suboptimal elsewhere. Yet the pressures of trillion-scale long-context MoE training are
sharply \emph{component-dependent}: attention is activation-bound, the MoE
feed-forward is weight- and routing-bound, the vocabulary projection produces the
single largest logit tensor, and the optimizer state dwarfs the weights themselves. No
one parallelism is Pareto-optimal across all of these---ZeRO-3 shards the expert
weights but leaves the full-rank MoE activations on every device, expert parallelism
shards experts but not the $\mathcal{O}(NkH)$ routed activations, and 3D parallelism
multiplies its axes against a fixed device budget while still applying the same
trade-off to every layer. The result is wasted headroom: at any moment most of the
machine's resources---idle CPU cores, host DRAM, and spare interconnect
bandwidth---sit unused while one bottleneck dictates the global plan.

\paragraph{Mixture-of-Parallelisms.} We propose \emph{Mixture-of-Parallelisms} (MoP), a
training stack that abandons the single global plan and instead \emph{specializes a
different parallelism to each component} of the MoE block, choosing for each one the
sharding axis that matches its dominant bottleneck (\S\ref{sec:method:plan}). The dense
path is sharded with an efficient sharding method, attention activations with sequence
parallelism \citep{jacobs2023ulysses}, expert weights with a hybrid
expert-parallel\,$\times$\,sharding scheme, the vocabulary with a sharded
data-tensor-parallel projection, and the optimizer is updated by an efficient pipeline
that exploits all of the machine's physical resources and leverages
computation--communication overlap. Crucially, every parallelism we compose is
\emph{data-parallel-like at the rank level}---each device holds a distinct shard of
tokens (under sequence parallelism, a different chunk of the same long sequence) and a
sharded copy of the weights it touches---so the components share \emph{overlapping
sub-groups of the same $W$ ranks} rather than forming a Cartesian product. This sidesteps
the rank multiplication of 3D parallelism while letting each component pick the sub-group
whose sharding axis dilutes its bottleneck. MoP is explicitly designed to exploit
\emph{all} of the machine's physical resources---GPU HBM, host DRAM,
and the CPU--GPU, GPU--GPU, and node--node bandwidths---rather than the GPUs alone.

To make this plan viable at trillion-parameter, million-token scale we build on three
lossless principles that attack the largest remaining peaks:
\begin{itemize}
    \item Building on Least-Loaded Expert Parallelism \citep{nguyen2026llep}, we adopt a \emph{memory-efficient expert-parallel
    scheme} that overlaps routing communication with expert computation to keep the transient
    MoE activation footprint small while preserving LLEP's speedup (\S\ref{sec:method:llep}).
    \item We shard the vocabulary projection with a \emph{data-tensor-parallel scheme} that
    computes the exact loss and gradient without ever materializing the $\mathcal{O}(NV)$
    logit tensor, while respecting the DP-like invariant that every rank carries a different
    batch (\S\ref{sec:method:ringtp}).
    \item We update the optimizer state
    through an \emph{efficient pipeline} that leverages a novel computation-communication overlap scheme. This collapses the optimizer time on the critical path (\S\ref{sec:method:streamadam}). 
\end{itemize} 
All three are \emph{lossless}: they relocate computation, not its result, preserving exact training semantics.

\paragraph{Results.} Together these components let MoP train models that previously
required far larger clusters on a fraction of the hardware. Across MoE models trained
from scratch at 120B, 600B, and 1T parameters---on 2, 8, and 12 $8{\times}$H200 nodes
respectively---MoP sustains $4.7\text{--}8.2\times$ higher per-GPU throughput than the
best-tuned FSDP2 baseline at the longest context that baseline can fit, with the
advantage widening as model scale grows (\S\ref{sec:eval}). More importantly, MoP keeps
training at context lengths up to $1$M tokens---where FSDP exhausts device memory beyond
$64\text{--}128$K---while holding throughput nearly flat as context grows, and its
per-step wall time is bound essentially by forward and backward compute even at
trillion-parameter scale. The combined memory savings enable \emph{lossless}
pre-training and fine-tuning of trillion-parameter MoE models at near-million-token
context on just under twelve $8{\times}$H200 nodes---a regime out of reach for prior
single-plan systems at the same hardware budget---by exploiting every tier of the
machine's compute, memory, and bandwidth.

\paragraph{Contributions.} In summary: \Ni we recast trillion-scale long-context MoE
training as a \emph{component-specialized} sharding problem and give a memory--bandwidth
bottleneck analysis that motivates a per-component parallelism assignment over
overlapping rank sub-groups (\S\ref{sec:method:taxonomy}--\ref{sec:method:plan}); \Nii we
contribute three lossless principles---a memory-efficient expert-parallel scheme, a
sharded data-tensor-parallel vocabulary projection, and an efficient optimizer
pipeline---that eliminate the dominant
activation and optimizer peaks (\S\ref{sec:method:llep}--\ref{sec:method:streamadam}); and
\Niii we show empirically that MoP delivers $4.7\text{--}8.2\times$ higher per-GPU
throughput than the best-tuned FSDP2 baseline and keeps training at up to
$1$M-token context where that baseline runs out of memory, enabling trillion-parameter,
million-context training on a minimal GPU footprint (\S\ref{sec:eval}).

\section{Background}\label{sec:background}

We review the memory model of large-model training that the rest of the paper
optimizes against (\S\ref{sec:bg:memory}), and the two parallelism families whose
trade-offs motivate our design: Megatron-style tensor/pipeline model parallelism
(\S\ref{sec:bg:megatron}) and ZeRO/FSDP sharded data parallelism
(\S\ref{sec:bg:fsdp}). Throughout, $\Theta$ denotes the number of trainable
parameters, $W$ the number of devices, $L$ the number of transformer blocks, $H$ the
model width, $S$ the sequence length, $B$ the per-rank micro-batch, and $N\!=\!BS$ the
per-rank tokens per forward pass.

\subsection{The Memory Model of Large-Model Training}\label{sec:bg:memory}

\paragraph{Mixture-of-Experts.} A Mixture-of-Experts (MoE) layer replaces the dense
feed-forward block with $E$ parallel expert networks and a lightweight router that
sends each token to its top-$k$ experts \citep{gshard_lepikhin2020,fedus2022switch}.
This decouples parameter count from per-token compute: total parameters grow with $E$
while the FLOPs per token grow only with $k\!\ll\!E$. Consequently, for the
trillion-parameter models we target, the overwhelming majority of $\Theta$ resides in
the routed expert weights, while each token activates only a small, \emph{data-dependent}
subset of them---a property that makes the memory and communication profile of MoE
training sharply non-uniform across layers and components.

\paragraph{Persistent vs.\ transient memory.} Per-device memory during training splits
into two regimes that must be reduced by different means. \emph{Persistent} (model)
state is resident for the whole step: under mixed-precision training with AdamW
\citep{loshchilov2019decoupled}, each parameter requires a $2$-byte (bf16) working copy,
a $2$-byte gradient, and $12$ bytes of fp32 optimizer state (a master weight plus the
first and second moments $m,v$), for
\begin{equation}
  M_{\mathrm{persist}} \;=\; \underbrace{2\Theta}_{\text{weights}} \;+\;
  \underbrace{2\Theta}_{\text{grads}} \;+\;
  \underbrace{12\Theta}_{\text{optimizer state}} \;=\; 16\Theta \ \text{bytes}.
  \label{eq:bg-persist}
\end{equation}
\emph{Transient} (activation) state is created in the forward pass and consumed in the
backward pass; it scales with the batch and sequence dimensions rather than with
$\Theta$, with a per-block peak of order $\mathcal{O}(NH)$ for the dense path and growing
with depth $L$ when activations are retained for backpropagation. Activation
checkpointing \citep{chen2016sublinear} trades compute for memory by discarding
intermediate activations in forward and recomputing them in backward. The central
difficulty of trillion-scale long-context MoE training is that no single technique
shrinks both regimes at once: sharding $\Theta$ attacks \eqref{eq:bg-persist} but leaves
activations untouched, while sequence/activation methods do the reverse.

\subsection{Tensor and Pipeline Parallelism (Megatron)}\label{sec:bg:megatron}

Megatron-LM \citep{shoeybi2019megatron} scales beyond a single device by partitioning
the \emph{model} across ranks. \emph{Tensor parallelism} (TP) over a group of $t$
devices splits each weight matrix along rows or columns so that every rank stores
$1/t$ of the dense weights and computes $1/t$ of each matmul; the partial results are
combined with two all-reduces per transformer block (one in attention, one in the
feed-forward). Because this communication sits on the critical path of every layer, TP
is restricted to within a high-bandwidth domain (e.g.\ NVLink inside a node).
\emph{Sequence parallelism} \citep{korthikanti2023reducing} complements TP by further
sharding the LayerNorm and dropout activations along the sequence axis, removing the
activation replication that pure TP leaves in those regions. \emph{Pipeline
parallelism} (PP) partitions the $L$ blocks into $p$ contiguous stages placed on
different devices and streams micro-batches through them; an idle ``bubble'' at pipeline
fill/drain costs throughput unless the micro-batch count is large.

\paragraph{Memory model.} Within a TP group, persistent state is reduced from
\eqref{eq:bg-persist} to $\approx\!16\Theta/t$ per rank and the corresponding
activations by up to $t$; within a PP group each stage holds only its $1/p$ of the
layers, i.e.\ $\approx\!16\Theta/p$ per stage. The three axes form a Cartesian
\emph{3D parallelism} with $W = d\,t\,p$, where $d$ is the data-parallel degree. This
factorization is the main drawback: the axes multiply to consume the device budget, $t$
and $p$ are bounded by intra-node bandwidth and the bubble respectively, and a single
global plan $(d,t,p)$ is applied uniformly to every layer regardless of whether a given
component is bottlenecked by weights, activations, or routing.

\subsection{Sharded Data Parallelism (ZeRO / FSDP)}\label{sec:bg:fsdp}

Plain data parallelism replicates the full $16\Theta$ bytes of \eqref{eq:bg-persist} on
every rank and synchronizes gradients with an all-reduce, which is memory-prohibitive at
scale. ZeRO \citep{rajbhandari2020zero} and its PyTorch realization FSDP
\citep{zhao2023fsdp} instead \emph{shard} the persistent state across the $W$
data-parallel ranks in three cumulative stages: ZeRO-1 shards the optimizer state
($2\Theta + 2\Theta + 12\Theta/W$ per rank), ZeRO-2 additionally shards gradients
($2\Theta + 14\Theta/W$), and ZeRO-3 / FSDP-\texttt{FULL\_SHARD} additionally shards the
parameters themselves, giving
\begin{equation}
  M_{\mathrm{persist}}^{\mathrm{ZeRO\text{-}3}} \;=\; 16\Theta / W \ \text{bytes per rank}.
  \label{eq:bg-zero3}
\end{equation}
Under ZeRO-3 each rank persistently owns a $1/W$ shard; just before a layer executes, an
all-gather reconstructs its full weights, the layer runs, and the gathered copy is
released, while gradients are reduce-scattered on the backward pass. The total
communication volume per step is comparable to a standard data-parallel all-reduce
(roughly $1.5\times$) but is spread across layers and overlappable with compute. ZeRO
keeps a key data-parallel property---each rank processes a \emph{different}
micro-batch---which makes it composable, but it shares data parallelism's blind spot:
it reduces only the persistent state of \eqref{eq:bg-persist} and leaves the full-rank
activations on every device. ZeRO-Offload \citep{ren2021zerooffload} pushes the optimizer
state and master weights (and optionally gradients) to host memory, trading PCIe/NVLink
host--device bandwidth for further device savings on the $12\Theta$-byte optimizer state.

\section{Mixture-of-Parallelisms}\label{sec:method}

\begin{figure}[t]
  \centering
  \includegraphics[width=\linewidth]{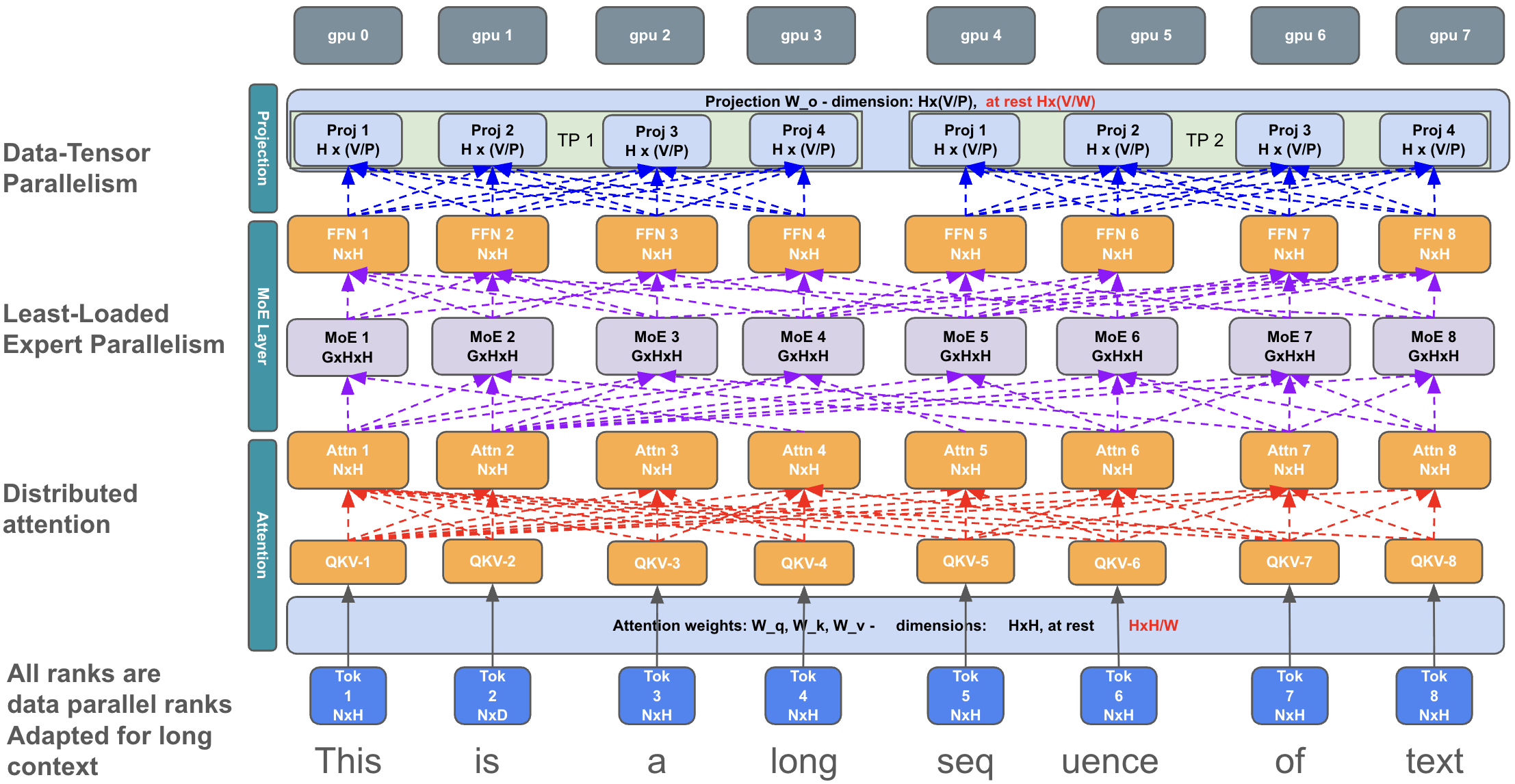}
  \caption{Overview of the Mixture-of-Parallelisms approach across 8 GPUs. Each layer
    component---attention (distributed attention), MoE experts (least-loaded expert
    parallelism), FFN, and projection (data-tensor parallelism)---uses a distinct
    parallelism strategy, all sharing the same $W$ data-parallel ranks without a
    Cartesian rank explosion.}
  \label{fig:mop-overview}
\end{figure}

We propose \emph{Mixture-of-Parallelisms} (MoP), a training stack for trillion-parameter
sparse Mixture-of-Experts (MoE) models at long context. Existing systems---tensor and
pipeline model parallelism \citep{shoeybi2019megatron}, fully-sharded data parallelism
\citep{rajbhandari2020zero,zhao2023fsdp}, expert parallelism
\citep{gshard_lepikhin2020,fedus2022switch,rajbhandari2022deepspeedmoe}, and sequence
parallelism \citep{korthikanti2023reducing,jacobs2023ulysses}---compose these techniques
along a single global plan applied uniformly to every layer. We argue that the memory
and bandwidth pressures of trillion-scale MoE training are sharply
\emph{component-dependent} and that a single parallelism is not Pareto-optimal across
attention, the dense path, the MoE feed-forward, the vocabulary projection, and the
optimizer. MoP composes a different parallelism for each component, attacking that
component's dominant bottleneck (\S\ref{sec:method:plan}); we describe a memory-efficient
expert-parallel scheme (\S\ref{sec:method:llep}), a sharded vocabulary projection
(\S\ref{sec:method:ringtp}), and an efficient pipeline for the optimizer state
(\S\ref{sec:method:streamadam}).

\subsection{Notation and Bottleneck Analysis}\label{sec:method:taxonomy}

Let $W$ be the number of devices, $L$ the number of transformer blocks, $H$ the model
width, $V$ the vocabulary size, $E$ the number of experts, $I$ the per-expert
intermediate width, $k$ the top-$k$ routing degree, $S$ the full sequence length, and
$N$ the number of tokens resident on each device per forward pass. We assume a single
sequence per step (batch size $B\!=\!1$); the sequence parallelism of \S\ref{sec:method:plan}
splits a length-$S$ sequence across a $D$-rank sub-group, so each device holds
$N\!=\!S/D$ tokens (equivalently $S\!=\!ND$), while distinct sub-groups hold distinct
sequences. Trainable parameters partition as
\begin{equation}
  \Theta \;=\; \underbrace{\Theta_{\mathrm{dns}}}_{\Theta(LH^{2})} \;+\;
              \underbrace{\Theta_{\mathrm{exp}}}_{\Theta(LEHI)} \;+\;
              \underbrace{\Theta_{\mathrm{voc}}}_{\Theta(HV)} ,
  \label{eq:param-decomp}
\end{equation}
covering dense weights (attention $Q,K,V,O$ projections, layer norms, MoE routers, the
embedding), expert weights (routed feed-forward), and the vocabulary projection. For
trillion-parameter MoEs $\sim\!98\text{--}99\%$ of $\Theta$ is in $\Theta_{\mathrm{exp}}$.

Per-rank GPU memory splits into two regimes that no single technique reduces
simultaneously: \textbf{persistent} parameter memory (master weights, gradients, optimizer
states \citep{loshchilov2019decoupled}), reduced by sharding or offloading
\citep{ren2021zerooffload}; and \textbf{transient} activation memory, scaling with $N$
and $H$, reduced by sequence sharding, activation checkpointing
\citep{chen2016sublinear,korthikanti2023reducing}, or operator-level chunking. The
dominant per-block activation peaks are
\begin{equation}
  M^{\mathrm{attn}}_{A} \;=\; \mathcal{O}(NH), \quad
  M^{\mathrm{moe}}_{A} \;=\; \mathcal{O}\!\bigl(NkH + N_{\mathrm{r}} I\bigr), \quad
  M^{\mathrm{voc}}_{A} \;=\; \mathcal{O}(NV),
  \label{eq:peaks}
\end{equation}
where $N_{\mathrm{r}}$ is the number of token--expert assignments received by the rank
under expert parallelism. Sharding parameters does not, by itself, attack activations:
ZeRO-3 reduces $\Theta_{\mathrm{exp}}$ to $\Theta_{\mathrm{exp}}/W$ but leaves
$M^{\mathrm{moe}}_{A}$ full-rank on every device; expert parallelism shards
$\Theta_{\mathrm{exp}}$ along $E$ but does not reduce
$\mathcal{O}(NkH)$. MoP exploits these asymmetries by treating each component
separately.

\subsection{Component-Specialized Parallelism Plan}\label{sec:method:plan}

\begin{table}[t]
  \centering
  \caption{Per-rank memory under MoP for one transformer-MoE block. $W$ is the total
    device count; $D$, $E_{p}$, $P$ are sub-group sizes for sequence, expert, and
    projection parallelism (with $D, E_{p}, P \le W$). Each row reports the regime it
    dominates.}
  \label{tab:mop-summary}
  \small
  \setlength{\tabcolsep}{5pt}
  \renewcommand{\arraystretch}{1.1}
  \begin{tabular}{@{}llcc@{}}
    \toprule
    \textbf{Component} & \textbf{Parallelism} &
    \textbf{Per-rank memory} & \textbf{Ref.} \\
    \midrule
    Dense weights $\Theta_{\mathrm{dns}}$ &
      Efficient sharding over $W$ &
      $\Theta_{\mathrm{dns}}/W$ &
      \S\ref{sec:method:plan} \\
    Attention activations &
      Sequence parallelism over $D$ &
      $\mathcal{O}(NH)$ &
      \S\ref{sec:method:plan} \\
    Expert weights $\Theta_{\mathrm{exp}}$ &
      Hybrid EP$\,\times\,$sharding (deg.\ $W$) &
      $\Theta_{\mathrm{exp}}/W$ &
      \S\ref{sec:method:plan} \\
    MoE activations &
      Memory-efficient LLEP \citep{nguyen2026llep} &
      $\ll\mathcal{O}(NkH)$ &
      \S\ref{sec:method:llep} \\
    Vocabulary weights $\Theta_{\mathrm{voc}}$ &
      Data-tensor parallelism over $P$ &
      $\Theta_{\mathrm{voc}}/P$ &
      \S\ref{sec:method:ringtp} \\
    Vocabulary logits &
      Sharded online loss &
      $\mathcal{O}(NV/P)$ &
      \S\ref{sec:method:ringtp} \\
    Optimizer state $\{m,v,\mathrm{master}\}$ &
      Host-resident; pipelined updates &
      small on-device working set &
      \S\ref{sec:method:streamadam} \\
    \bottomrule
  \end{tabular}
\end{table}

Table~\ref{tab:mop-summary} summarizes the assignment. A unifying property is that, in
contrast to pipeline or tensor parallelism, every parallelism we compose is
\emph{data-parallel-like at the rank level}: each device holds a distinct shard of
tokens---under sequence parallelism, a different chunk of the same long sequence---and
stores a (sharded) copy of the weights involved. The $W$ ranks form
overlapping sub-groups of sizes $D$, $E_{p}$, and $P$ along which different components
shard; these sub-groups are not independent dimensions of a Cartesian product but
subsets of the same $W$ ranks. This avoids the rank multiplication
$W=D_{p}T_{p}P_{p}$ of classical 3D parallelism while still letting each component
pick the sub-group whose sharding axis matches its bottleneck.

\paragraph{Dense weights via efficient sharding.} The dense path
($\Theta_{\mathrm{dns}}\!\ll\!\Theta_{\mathrm{exp}}$) holds the attention $Q,K,V,O$
linears, layer norms, MoE routers, and the input embedding. Dense weights are touched by
every token, so they cannot be sharded along the token axis without communication on
every layer. We shard them along the parameter axis with an efficient sharding method:
each rank persistently stores
$\Theta_{\mathrm{dns}}/W$ parameters; just before a layer fires, an
all-gather over the $W$-rank sub-group reconstructs the full weight matrix on every rank,
the layer is computed, and the gathered copy is released. Gradients follow the inverse
schedule via reduce-scatter. The all-gather volume per layer is
$\mathcal{O}(\Theta_{\mathrm{dns}}^{(\mathrm{layer})}\,W)$, which is small in absolute
terms because $\Theta_{\mathrm{dns}}$ itself is small (1--2\% of $\Theta$); for the
trillion-parameter MoEs we target, this is sub-percent of step time at $W\!=\!10^{2}$.

\paragraph{Attention activations via sequence parallelism.} At long context the
attention block dominates forward memory. A length-$S$ sequence induces an
$\mathcal{O}(S^{2})$ attention-score matrix; the IO-aware FlashAttention kernel
\citep{dao2022flashattention} never materializes it, but the activations it streams
over (the $Q,K,V$ projections and the attention output) still cost $\mathcal{O}(SH)$
for the full sequence. We distribute this across a $D$-rank sub-group by sharding the
sequence, following Ulysses-style sequence parallelism
\citep{jacobs2023ulysses,korthikanti2023reducing}, so that each rank holds only
$N\!=\!S/D$ tokens and the full-sequence cost $\mathcal{O}(SH)$ becomes a per-rank
$\mathcal{O}(NH)$. Outside the attention block, all tensors live in a
\emph{sequence-sharded} layout, with each rank holding its $N\!=\!S/D$-token chunk of
the sequence across the full hidden width $H$. Inside the attention block, an
all-to-all swaps to a \emph{head-sharded} layout, with each rank holding the full
$S\!=\!ND$ tokens across an $H/D$ slice of the hidden width (a $1/D$-th of the heads),
so that the row-wise softmax has access to the entire sequence; a second all-to-all
reverses the swap on exit. Either layout exposes the same per-rank activation peak
$\mathcal{O}(NH)$ ($=\mathcal{O}(SH/D)$). On top of this scheme we adopt more advanced
upgrades to Ulysses-style sequence parallelism that further lower its memory footprint, enabling longer context training.
Critically, $D$ is chosen independently of the expert-parallel degree
(\S\ref{sec:method:plan}), so we can grow $D$ to dilute attention activations without
disturbing the expert sharding plan.

\paragraph{Expert weights via hybrid EP$\,\times\,$sharding.} The MoE feed-forward holds
$\Theta_{\mathrm{exp}}$, essentially the entire parameter mass of the model. Pure
exclusive expert parallelism \citep{gshard_lepikhin2020,fedus2022switch} assigns $E/W$
experts to each rank and dispatches tokens to their destination expert by an all-to-all,
but it couples the expert-parallel degree to the total device count and inflates the
routing volume of every layer. We instead compose two sharding axes---an expert-parallel
sub-group over which experts are partitioned and a complementary sharding sub-group over
which each expert shard is further sharded---so that
persistent per-rank expert memory is $\Theta_{\mathrm{exp}}/W$, while the split between
the two axes trades inter-rank routing volume against weight-gather volume. The MoE
activation peak is then attacked separately by the memory-efficient LLEP scheme of
\S\ref{sec:method:llep}.

The remaining three components---MoE activations, the vocabulary projection, and the
optimizer state---introduce new mechanisms described next.

\subsection{Memory-Efficient Expert Parallelism}\label{sec:method:llep}

\paragraph{Recap of LLEP.} \citet{nguyen2026llep} observed that even well-trained MoE
models exhibit highly imbalanced expert routing during fine-tuning and inference, which
violates standard expert parallelism's assumption of uniform load and concentrates
work on the devices hosting popular experts---incurring stragglers and out-of-memory
failures. \emph{Least-Loaded Expert Parallelism} (LLEP) addresses this at the system
layer: at every MoE call it measures per-rank token load, migrates excess tokens from
overloaded ranks to underloaded ones, and ships the corresponding expert parameters
along with the migrated tokens so that the receiving rank can compute the expert
feed-forward locally. LLEP preserves model semantics (it permutes the placement of
compute, not the assignment of tokens to experts), supports backward gradient
propagation, and delivers up to $5\times$ MoE-layer speedup and $4\times$ peak-memory
reduction over standard expert parallelism. We adopt LLEP unchanged as our routing layer.

\paragraph{A memory-efficient variant.} LLEP balances \emph{which} device runs each
token--expert assignment, but the per-rank \emph{transient} forward peak of a MoE layer
still scales with the rank's incoming token count, and the routed inputs, the dispatch
buffers, and the expert intermediate are otherwise live at the same time. On top of LLEP
we adopt a memory-efficient variant that overlaps the routing communication with the
expert computation, so that only a small fraction of these buffers is resident at any
instant. The scheme substantially lowers the dominant transient MoE activation peak
while recovering LLEP's throughput, enabling longer context or larger micro-batches at
the same hardware budget. It is lossless: it reorders and overlaps computation and
communication without changing the layer's result.

\subsection{Data-Tensor-Parallel Vocabulary Projection}\label{sec:method:ringtp}

The vocabulary projection forms the largest single weight matrix in modern LLMs
($V\!\gtrsim\!10^{5}$) and the largest activation tensor when $S$ is large. With local
input batch $X\!\in\!\mathbb{R}^{N\times H}$ and projection weight
$W_{\mathrm{voc}}\!\in\!\mathbb{R}^{H\times V}$, the forward pass computes
\begin{equation}
  Y \;=\; X\,W_{\mathrm{voc}} \;\in\; \mathbb{R}^{N\times V},
  \qquad
  \ell_{n} \;=\; Y_{n,t_{n}} \;-\; \log\!\sum_{v=1}^{V} \exp\!\bigl(Y_{n,v}\bigr),
  \label{eq:voc-fwd-loss}
\end{equation}
where $t_{n}$ is the target vocabulary index. Storing $Y$ alone is $\mathcal{O}(NV)$,
exceeding device memory for long sequences ($\sim\!10^{2}$\,GB at $N\!=\!10^{6}$,
$V\!=\!2{\times}10^{5}$). Standard column-wise tensor parallelism shards
$W_{\mathrm{voc}}$ across $P$ ranks but \emph{requires every rank to hold the same
$X$}, incompatible with our DP-like plan in which every rank holds a different batch.

\paragraph{A sharded data-tensor-parallel projection.} We shard $W_{\mathrm{voc}}$
across a $P$-rank sub-group while preserving the DP-like invariant that each rank keeps
its own distinct batch. The scheme computes the exact per-token loss and its gradients by
combining per-block partial contributions, so that the full $\mathcal{O}(NV)$ logit
tensor $Y$ is never materialized in either the forward or backward pass---only a fraction
of it is resident at any instant. Per-rank peak activation memory for the projection is
thereby held at $\mathcal{O}(NV/P)$ rather than $\mathcal{O}(NV)$, an $8\text{--}16\times$
reduction in practice at the sub-group sizes we use, at a communication cost that---for
large vocabularies---is essentially independent of context length. The result is
lossless: the loss and gradients match the dense projection to numerical precision.

\subsection{Efficient Optimizer Pipeline}\label{sec:method:streamadam}

For trillion-parameter MoEs, the optimizer state---for AdamW, the master weights and two
moments $(m,v)$---contributes $\sim\!12\Theta$ bytes in fp32, an order of magnitude larger
than the bf16 weights themselves \citep{loshchilov2019decoupled}, so keeping it resident in
device memory alongside long-context activations is infeasible. Rather than reserve scarce
GPU memory for it, MoP exploits \emph{all} of the machine's physical resources---not the
GPUs alone---and drives the optimizer step through an efficient pipeline that leverages
computation--communication overlap: the update is issued in parallel with work already on
the critical path (including the backward pass) and uses only a small transient device
working set, so its cost is effectively removed from the per-step wall time. Combined with
the components above, this leaves the end-to-end step time bound essentially by forward and
backward compute, even at trillion-parameter scale. The update is lossless: it reorders
data movement and computation without altering the optimizer's result.


\begin{table}[t]
  \centering
  \caption{Training throughput (tokens/s/GPU) of MoP vs.\ the best-performing FSDP2
    baseline (FSDP-best) for MoE models trained from scratch on $8\!\times\!$H200 nodes.
    FSDP-best is the highest-throughput configuration of a tuned FSDP2/TorchTitan stack
    \citep{zhao2023fsdp,liang2025torchtitan} that also supports Megatron-style TP/CP/EP
    and CPU offloading (per-model configurations in the text). ``OOM'' marks
    configurations where the baseline runs out of device memory; the speedup is reported
    at the longest context FSDP can fit.}
  \label{tab:throughput}
  \small
  \setlength{\tabcolsep}{10pt}
  \renewcommand{\arraystretch}{1.15}
  \begin{tabular}{@{}llrrrr@{}}
    \toprule
    \textbf{Model} & \textbf{Nodes} & \textbf{Context} &
    \textbf{MoP} & \textbf{FSDP-best} & \textbf{Speedup} \\
    \midrule
    \multirow{2}{*}{120B} & \multirow{2}{*}{2}  & 128K & 1043    & 223 & $4.7\times$ \\ 
                          &                     & 256K & 681     & OOM & --- \\
                          &                     & 512K & 424     & OOM & --- \\
                          &                     & 1M   & 393     & OOM & --- \\
    \midrule
    \multirow{5}{*}{600B} & \multirow{5}{*}{8}  & 64K  & 236 & 39  & $6.1\times$ \\ 
                          &                     & 128K & 220 & OOM & --- \\
                          &                     & 256K & 225 & OOM & --- \\
                          &                     & 512K & 213 & OOM & --- \\
                          &                     & 1M   & 213 & OOM & --- \\
    \midrule
    \multirow{2}{*}{1T}   & \multirow{2}{*}{12} & 128K & 166 & 25  & $6.6\times$ \\ 
                          &                     & 256K & 140 & 17  & $8.2\times$ \\ 
                          &                     & 512K & 108 & OOM & --- \\
                          &                     & 1M   & 100 & OOM & --- \\
    \bottomrule
  \end{tabular}
\end{table}

\section{Experiments}\label{sec:eval}

\begin{table}[t]
  \centering
  \caption{Best-performing FSDP2 baseline (FSDP-best) parallelism configuration per model,
    obtained by sweeping data (FSDP), expert (EP), context (CP), and tensor (TP)
    parallelism together with optimizer CPU offloading. These are the configurations used
    for the FSDP-best entries in Table~\ref{tab:throughput}.}
  \label{tab:fsdpconfig}
  \small
  \setlength{\tabcolsep}{10pt}
  \renewcommand{\arraystretch}{1.15}
  \begin{tabular}{@{}lccccc@{}}
    \toprule
    \textbf{Model} & \textbf{FSDP} & \textbf{EP} & \textbf{CP} & \textbf{TP} &
    \textbf{CPU offload} \\
    \midrule
    120B & 1  & 16 & 16 & --- & No  \\
    600B & 16 & 8  & 2  & 2   & Yes \\
    1T   & 3  & 96 & 4  & 8   & Yes \\
    \bottomrule
  \end{tabular}
\end{table}

\paragraph{Baselines.} We compare against a strong, fully featured FSDP2 baseline
\citep{zhao2023fsdp,liang2025torchtitan} as implemented in TorchTitan
\citep{liang2025torchtitan}, which is not limited to plain data-parallel sharding: it
also supports Megatron-style tensor (TP), context (CP), and expert (EP) parallelism
\citep{shoeybi2019megatron,korthikanti2023reducing}, as well as CPU offloading of the
optimizer state \citep{ren2021zerooffload}. For each model we sweep these axes and report
the highest-throughput configuration that fits in device memory (\emph{FSDP-best}); the
resulting per-model configurations are listed in Table~\ref{tab:fsdpconfig}. Our
baseline therefore already incorporates the best of existing parallelism strategies, so
the gains in Table~\ref{tab:throughput} are over this tuned combination rather than
vanilla FSDP.

\paragraph{Throughput and long-context scaling.} Table~\ref{tab:throughput} compares
MoP against the best-tuned FSDP2 baseline (FSDP-best) on MoE models trained from scratch,
each on $8\!\times\!$H200 nodes. At the longest context the FSDP baseline can fit, MoP
delivers $4.7\text{--}8.2\times$ higher per-GPU throughput, and the gap widens with model
scale. More importantly, MoP keeps training at context lengths up to $1$M tokens---where
FSDP exhausts device memory beyond $64\text{--}128$K---reaching configurations that are
simply infeasible for the baseline while holding throughput nearly flat as context grows.

\paragraph{Limitations.} MoP trades memory for communication: its heavy reliance on
parameter sharding (\S\ref{sec:method:plan}) and all-to-all routing (\S\ref{sec:method:llep}) makes
it \emph{communication-bound}, so the speedups in Table~\ref{tab:throughput} assume the
fast interconnect of our reference cluster. They erode when bandwidth is throttled or the
device count grows large enough that these collectives stop overlapping with compute. In
those regimes pipeline parallelism (which keeps traffic mostly intra-node) and an extra
fully separated data-parallel dimension (which replicates rather than shards state) help
recover throughput at some memory cost. MoP is thus the memory-efficiency extreme of the
design space, preferred when device memory---not bandwidth---is the binding constraint, as
in post-training of large MoE models on a small hardware budget.

\section{Conclusion}

We presented \emph{Mixture-of-Parallelisms} (MoP), a training stack that replaces the
single global parallelism plan with a \emph{component-specialized} assignment: each part
of the MoE block---the dense path, attention, the expert feed-forward, the vocabulary
projection, and the optimizer state---is sharded along the axis that matches its
dominant bottleneck, over overlapping sub-groups of the same $W$ ranks rather than a
Cartesian product. To make this plan viable at trillion-parameter, million-token scale,
we build on three lossless principles---a memory-efficient expert-parallel scheme, a
sharded data-tensor-parallel vocabulary projection, and an efficient pipeline for
host-offloaded optimizer state---that eliminate the largest transient activation and
optimizer peaks while preserving exact training semantics.
Empirically, MoP sustains $4.7\text{--}8.2\times$ higher per-GPU throughput than the
best-tuned FSDP2 baseline and keeps training at context lengths up to $1$M tokens
where that baseline runs out of memory, enabling lossless trillion-parameter training at
near-million-token context on just under twelve $8{\times}$H200 nodes. More broadly, our
results suggest that matching parallelism to component---rather than applying one plan
uniformly to every layer---is a practical route to training ever-larger sparse models on
commodity-scale clusters. Promising directions include automating the per-component
sub-group sizing ($D$, $E_{p}$, $P$) from the model shape and cluster topology, and
extending the component-specialized view to long-context inference.

\bibliography{neurips_2023.bib}
\bibliographystyle{plainnat}


\end{document}